\documentclass[twoside,twocolumn,9pt]{article}
\usepackage{extsizes}                         
\usepackage[left=2cm,right=2cm,top=1.785cm,bottom=2.0cm]{geometry}
\usepackage[T1]{fontenc}
\usepackage{mathptmx}                          
\usepackage{droidsans}                         
\usepackage[super,sort&compress,comma]{natbib} 
\usepackage{graphicx}
\usepackage{amsmath,amssymb,bm}
\usepackage{siunitx}
\usepackage[usenames,dvipsnames]{xcolor}
\usepackage{balance}                           
\usepackage[format=plain,justification=justified,singlelinecheck=false,
            font={stretch=1.125,small,sf},labelfont=bf,labelsep=space]{caption}
\usepackage[compact]{titlesec}

\usepackage{microtype}      
\titleformat*{\section}{\normalsize\bfseries\sffamily\raggedright}
\titleformat*{\subsection}{\small\bfseries\sffamily\raggedright}
\titleformat*{\subsubsection}{\small\itshape\sffamily\raggedright}
\usepackage{hyperref}
\hypersetup{colorlinks=true,linkcolor=blue,citecolor=blue,urlcolor=blue}
\begin{document}

\twocolumn[{%
  \begin{@twocolumnfalse}
  \begin{center}
    {\LARGE\bfseries\sffamily Mapping bimodal colloidal depletion systems onto the\\ monodisperse gelation phase diagram\par}
    \vspace{1.5ex}
    {\large Robert A. Campbell\textsuperscript{1}, Ziye Zhuang\textsuperscript{2},
            Ali Mohraz\textsuperscript{2}, Safa Jamali \textsuperscript{1}\par}
    \vspace{1ex}
    {\itshape\small
      \textsuperscript{1}Department of Mechanical and Industrial Engineering,
        Northeastern University, Boston, MA 02115, USA\\
      \textsuperscript{2}Department of Chemical and Biomolecular Engineering,
        University of California, Irvine, CA 92697, USA\par}
  \end{center}
  \vspace{2ex}
  \vspace{1.5ex}
  {\noindent\textbf{Abstract}\quad Many industries rely on models of colloidal gelation and phase behavior to predict how processing equipment will handle new product formulations. In colloidal systems with nearly uniform particle size it is possible to predict the formation of a gel from only three state variables; however, the extension of this model to bimodal and multimodal systems is not well described. Here we present a systematic simulation study of the attractive bimodal depletion gel phase diagram. For these systems, we find that gelation is highly dependent on total volume fraction, attraction conditions, and bimodal composition. Attraction and composition can be effectively collapsed into a single parameter using the contact-weighted second virial coefficient. This successfully projects the bimodal phase diagram onto the established monodisperse phase diagram, including accurately describing most bulk assembly properties; however, the underlying network structure of these gels remains highly dependent on the details of bimodal composition and attraction.\par}
  \vspace{1.5ex}
  \vspace{3ex}
  \end{@twocolumnfalse}
}]


\section*{Introduction}

Colloidal gels are nonequilibrium soft solids formed through the aggregation and dynamical arrest of attractive particles. Their multiscale structures give rise to tunable mechanical and transport properties that are important in foods, personal-care formulations, tissue-engineering materials, cementitious systems, and battery slurries \cite{Mezzenga2005,Diba2021,Duduta2011}. For approximately monodisperse particles interacting through short-range, isotropic attractions, gelation can be rationalized using three primary state variables: particle volume fraction, attraction strength, and attraction range \cite{Pusey1986,Lu2006-fluidsofclusters,Lu2008,Lu2013}. This corresponding-state description provides a useful framework for predicting whether a suspension remains dispersed, forms finite clusters, or develops a persistent space-spanning network. One of the biggest strengths of this model is that these non-dimensional parameters are system neutral. Although these state-dependent relationships were largely discovered in depletion gels, where micron sized particles interact through the volumetric exclusion of a smaller polymer depletant, the same phase behavior holds for any Brownian particulate system with short range, attractive, centro-symmetric interactions. As such, these models are regularly used in systems other than depletion gels, such as proteins\cite{Stradner2020} and cement hydrates\cite{Ioannidou2016}.

Most practical colloidal formulations, however, are not monodisperse. Mixtures containing multiple particle sizes introduce several coupled effects that are absent from the monodisperse framework \cite{Royall2021}. At fixed total volume fraction, composition changes the particle-number ratio, collision frequency, diffusivity, contact geometry, local coordination capacity, and available packing configurations. In depletion suspensions, size dispersity also generates pair-dependent interactions because the overlap volume excluded to the depletant differs among small–small, small–large, and large–large particle pairs \cite{Asakura1958,Vrij1976}. It is therefore not evident whether the effects of composition can be represented by a single effective attraction strength or whether each pair interaction must be treated independently.

Previous studies have shown that particle size, interaction asymmetry, and the addition of secondary particle populations can alter colloidal assembly in several ways \cite{Jiang2025,LonialWeeks2026,FerreiroCordova2020,DiMichele2014,Yuan2016,Sikorski2011,LiRoyerSunNess2023,JiangSeto2023,FerreiroCordova2022,Mohraz2008,Jia2015,Jia2018}. Non-attractive inclusions may modify the available free volume, disrupt the gel backbone, or introduce a second structural length scale \cite{Jiang2025,LiRoyerSunNess2023,JiangSeto2023,FerreiroCordova2022,Mohraz2008,Jia2015,Jia2018}. Selectively interacting binary colloids can exhibit composition-dependent percolation sequences and aggregation times \cite{DiMichele2014}, while highly size-asymmetric depletion mixtures can develop population-dependent diffusion and demixing \cite{Sikorski2011}. Depletion based bimodal colloids have primarily been studied through the addition of non-interactive inclusions with high particle size asymmetries. The heterogeneous dynamics in these systems potentially alter aggregation kinetics, changing both the gel line and resulting gel structure. In the gel regime, recent studies have confirmed that the addition of large non-Brownian granular inclusions alters the phase boundaries of the gel transition in two different ways, depending on filler size and gel structure. In the "passive" case, these inclusions fit into existing voids in the network and do not significantly alter gel formation. In the "active" case they disrupt gel structure by introducing new voids, decreasing the available free volume and effectively increasing the particle volume fraction\cite{LiRoyerSunNess2023}. These filler effects appear to be highly size dependent. At high total volume fractions ($\phi=0.6$) the mechanical effect of large inclusions can be well described by continuum models; however, as the size ratio approaches 1:1 there are greater disruptions to the original network, and these disruptions can mechanically enhance the system beyond the continuum prediction\cite{FerreiroCordova2022}. This mirrors results from lower volume fractions, where mixtures of similarly sized attractive and repulsive particles will trap repulsive inclusions, arresting their dynamics and producing a more spatially homogeneous attractive network with fewer large voids and more tenuous contacts\cite{Mohraz2008}. Increasing the fraction of repulsive inclusions can ultimately push a gel network into the glass regime\cite{Jia2015}. In addition to increasing the effective volume fraction, these inclusions alter the localization length of local structural features, physically restructuring the system and altering its mechanics. In gels, it has been reported that an increased localization length mechanically weakens the network by reducing the density of interparticle bonds, but in glasses this strengthens the material by increasing caging effects\cite{Jia2018}. This suggests that bimodality may induce competing multiscale effects that alter both phase behavior and bulk mechanics. Together, these results demonstrate that multicomponent suspensions cannot generally be understood by replacing all particles with a single average size. However, a corresponding-state parameter capable of organizing the gelation boundary of fully attractive, size-bidisperse suspensions has not been established.

The reduced second virial coefficient provides a natural starting point for such a description because it integrates the strength and range of a pair potential into a single thermodynamic quantity \cite{NoroFrenkel2000}. For a mixture, however, different pair types possess different virial coefficients and occur with different statistical and geometric weights. A successful effective parameter must therefore account for both the interaction matrix and the mixture composition. Moreover, even if such a parameter predicts the formation of a percolated network, it may not predict the network’s species-resolved topology or multiscale structure. Distinguishing state equivalence from structural equivalence is particularly important because gels with similar average coordination or connectivity can display substantially different mechanical responses \cite{Nabizadeh2024,Rocklin2021,Whitaker2019}.

Here, we use core-modified dissipative particle dynamics simulations to study attractive bidisperse suspensions with a particle-radius ratio $a_L/a_S=2$. We vary the total particle volume fraction, the mixture composition, and the base attraction strength while maintaining a fixed short interaction range. We first determine how composition shifts the finite-time transition from dispersed particles to finite clusters and persistent percolated networks. We then test whether pairwise second virial coefficients can be combined into an effective composition-dependent interaction parameter that maps the bidisperse state diagram onto a monodisperse reference diagram. Finally, we examine where this mean-field mapping fails by resolving aggregation dynamics, coordination, mobility, local packing, and pore structure by particle population. We find that the effective attractive parameter organizes most bulk state classifications, whereas the resulting network structure remains strongly dependent on composition and on the underlying hierarchy of pair interactions.

\section*{Methods}

\subsection*{Bimodal depletion interactions}

Depletion interactions emerge from the volumetric exclusion of a small depletant particle between two larger colloids. In these systems the attraction strength is physically set by the depletant concentration, relative to the overlap concentration, and the attraction range is typically set by the radius of gyration of this depletant. However, in bimodal systems, a larger colloid radius will produce a larger effective overlap volume, increasing the effective attraction for these particles when compared to smaller colloids in the same system. In this case, the same depletant concentration and depletant size will produce heterogeneous interaction conditions between small-small, small-large, and large-large colloid pairs. We do not simulate explicit depletant particles in our systems. Instead, we represent this size-dependent change in volumetric exclusion effects through a size-weighted Morse potential. We scale the strength of attraction at particle contact with the average size of each interacting colloid pair, such that $U_0^{ij}=D_0(a_{i}+a_{j})/2$. At particle size ratio 1:2, this approximation produces attraction strengths in agreement with the Asakura-Oosawa-Vrij model for depletion systems \cite{Asakura1958,Vrij1976}. Pair-wise interactions are therefore calculated as:

\begin{equation}
    \mathbf{U}_{ij}^{Morse} (r_{ij}) = U^{ij}_0 \big(e^{-2\alpha (r_{ij}-(a_i+a_j))}- 2e^{-\alpha (r_{ij}-(a_i+a_j))}\big)
    \label{eq:morse-ch4}
\end{equation}
\\
\noindent
where $a_i$ and $a_j$ are the particle radii, $r_{ij}$ is the center-center separation distance and $\alpha$ is the attraction range parameter. In this study small particles are defined with $a_S=1$ and large particles are defined with $a_L=2$. All systems are simulated at a fixed attraction range representing the same depletant with $\Delta = r_g/a_S = 0.05$ and $\alpha=60/a_S$. Unless otherwise noted, we report attraction strength as the base attraction parameter for each system, $D_0$, which represents the effective interaction between two small-small particles of $a_S=1$. In the bimodal systems, there is always a corresponding attraction hierarchy, such that $U_0^{SS}=D_0$, $U_0^{SL}=1.5D_0$ and $U_0^{LL}=2D_0$. 

\subsection*{DPD Simulations}
Colloidal gels were simulated using a modified Dissipative Particle Dynamics (DPD) method \cite{Whittle2010,Jamali2015,Boromand2017}. This is implemented in a custom version of the open-source molecular dynamics toolkit HOOMD-blue \cite{Anderson2020}. DPD uses an explicit solvent, which ensures that both short-range and long-range hydrodynamic interactions are included. Integration is performed at constant number of particles, constant volume, and constant temperature (NVT), where the system temperature $kT=0.1$ is maintained through a fluctuation-dissipation relation. A two-step velocity-Verlet algorithm is used to ensure forces progress smoothly between discrete timesteps. The equation of motion for particle $i$ with mass $m_i$ and velocity $\mathbf{v}_i$ is described by six pair-wise forces:
\begin{equation}
m_{i} \frac{d\mathbf{v}_i}{dt} = \sum_{i,i \neq j}^{N}\big( \mathbf{F}_{ij}^{Cons} + \mathbf{F}_{ij}^D + \mathbf{F}_{ij}^R + \mathbf{F}_{ij}^{M} + \mathbf{F}_{ij}^H + \mathbf{F}_{ij}^{Cont}\big)
\label{eq:DPD-colloid-ch4}
\end{equation}
\noindent
In this framework, the solvent-solvent and solvent-colloid interactions are described by the first three force terms, which represent standard DPD interactions. These are a soft conservative force, $F_{ij}^{C}$ , the dissipative force, $F_{ij}^D$, and the random force, $F_{ij}^R$, defined as:
\begin{equation}
\mathbf{F}_{ij}^{Cons} = a_{ij} \omega_{ij}(r_{ij}) \mathbf{e}_{ij}
\label{eq:DPD-Conservative-ch4}
\end{equation}
\begin{equation}
\mathbf{F}_{ij}^D = -\gamma_{ij} \big[\omega_{ij}(r_{ij})\big]^2 \big( \mathbf{v}_{ij} \cdot \mathbf{e}_{ij} \big) \mathbf{e}_{ij}
\label{eq:DPD-Dissipative}
\end{equation}
\begin{equation}
\mathbf{F}_{ij}^R = \sigma_{ij} \omega_{ij}(r_{ij}) \Theta_{ij} \Delta t^{-2}\mathbf{e}_{ij}
\label{eq:DPD-Random-ch4}
\end{equation}
\noindent
The controlling parameter for the conservative force is the maximum attraction strength for the background fluid, $a_{ij}=25kT$, as calculated from the compressibility and density of water. The pair-specific weight term, $\omega_{ij}(r_{ij})=(1-(r_{ij}/r_{cut}))$, is calculated from the separation distance $r_{ij}$ and a maximum neighbor cut-off distance, $r_{cut}$, and the $\mathbf{e}_{ij}$ parameter is the directional vector of the contact. The dissipative and random forces are coupled through the fluctuation-dissipation relation such that their controlling parameters are defined in terms of each other: $\sigma_{ij}=\sqrt{2\gamma_{ij}kT}$. Additionally, the viscous resistance parameter, $\gamma_{ij}$, scales with the background viscosity, $\eta_0$, here set as $\gamma_{ij}=4.5, \eta_0=0.3$ to ensure high resolution of particle motion. The theta parameter, $\Theta_{ij}$, is a random-number white noise parameter, and $\Delta t$ is the integration timestep size. 

In colloid-colloid interactions $F_{ij}^{C}=0$ and interactions are defined by the last five force terms. This includes the dissipative and random forces as well as the Morse attractive force $F_{ij}^{M}$, a short-range lubrication term, $\mathbf{F}_{ij}^{H}$, and a hard-sphere repulsive contact force, $\mathbf{F}_{ij}^{Cont}$. These forces are defined in terms of the in terms of the surface-surface particle separation distance, $h_{ij}=r_{ij}-(a_i+a_j)$, as:
\begin{equation}
\mathbf{F}_{ij}^{M} = \frac{d\mathbf{U_{ij}^M}}{d\mathbf{r}_{ij}} = -2 \alpha D_0 \big( e^{-2 \alpha h_{ij}} - e^{- \alpha h_{ij}}\big)
\label{eq:DPD-MorseForce-ch4}
\end{equation}
\begin{equation}
\mathbf{F}_{ij}^{Cont} = 
    \begin{cases}
      f_{cont}(1+h_{ij}) &  h_{ij} \leq 0,\\
      f_{cont}(1-\frac{h_{ij}}{\delta_C}) &  0 \leq h_{ij} \leq \delta_C,\\
      0 &  h_{ij} > \delta_C,
    \end{cases}  
\label{eq:DPD-Contact-ch4}
\end{equation}
\begin{equation}
\mathbf{F}_{ij}^H = \mathbf{F}_{ij}^{sq} = -/mu_{ij}^H\big( \mathbf{v}_{ij} \cdot \mathbf{e}_{ij} \big)\mathbf{e}_ij
\label{eq:DPD-Hydro-ch4}
\end{equation}
\noindent
The contact force is defined by a magnitude $f_{cont}/kT=100$ and activation range of $\delta_C=0.001$ from the particle surface. The hydrodynamic lubrication force is calculated from the squeezing-coefficient for two particles interacting as a coordinated dumbell at low Reynolds number\cite{BallMelrose1997}. This is the pair-drag term:
\begin{equation}
\mu_{ij}^{H} = \frac{3 \pi \eta_0}{8 h_{ij}}\bigg( \frac{4a_ia_j}{a_i+a_j} \bigg)^2
\label{eq:DPD-a_squeeze-ch4}
\end{equation}

All particles are density matched so $m_S=1$ and that $m_C = 4/3 \rho \pi a_C^3$, where $\rho=3$ is the solvent number density. Systems are initialized in a cubic box of side length $L= 70a_S$ with periodic boundary conditions. Gelation behavior is studied across six total volume fractions, $\phi=[0.05, 0.10, 0.15, 0.20,0.25, 0.30]$, five large particle fractions $x_L=\phi_L/\phi=[0.0, 0.25, 0.5, 0.75, 1.0]$, and five attraction conditions $D_0/kT=[3,5,7,9,12]$. At constant box size, these systems vary in particle count, ranging from 512 to 24,566 colloidal particles. Each bimodal composition contains at least 500 particles of each population. In addition, each system contains between 720,299 and 926,100 solvent particles to carry long range hydrodynamic forces.

The simulations were performed in two steps. In the first step, all particles were randomly placed in the simulation box and allowed to briefly equilibrate without colloid-colloid attraction. This allows the simulation to resolve non-physical overlaps and fully distribute particles under quiescent conditions. In the second step, attractive interactions are added and the system is allowed to evolve for up to $180\tau^{D}_{\langle a_C \rangle}$, defined by the bare particle diffusion time of the average sized particle in the system $\tau^D_{\langle a_C \rangle}= 6\pi\eta_0 a_C^3/ k{T}$. Systems near the phase boundary were run the longest to clarify phase behavior.

\subsection*{Phase classification}

We classify each system as either a monomeric fluid, a fluid of clusters, or a gel, following the classification schema of the monodisperse colloidal phase diagram introduced by Lu et al.\cite{Lu2006-fluidsofclusters}. These three phases are defined using purely structural and dynamical definitions, consistent with previous structure-based studies of colloidal gels\cite{PanizHaghighi2025,Nabizadeh2024}. Many of our systems exist near the gelation boundary and could be classified as a fourth marginal phase; however, for greater visual clarity in the phase diagram we assign all of them to one of the three core phases, using the measures described below. 

In this phase classification scheme, a monomeric fluid is defined as a system with diffusive dynamics and average coordination number $\langle Z \rangle \leq 1$. This phase is separated from a fluid of clusters by the dimerization convention, where a system of clusters is defined by having $\langle Z \rangle > 1$. Structurally, a gel is differentiated from a fluid of clusters by the formation of a percolated, space-spanning network. The strictest definition of a gel also includes the emergence of a mechanical elastic modulus from this structure. Conventionally, the existence of a mechanically elastic network can be estimated from the average coordination number. In adhesive systems it is possible to form a mechanically load-bearing network at as low as $\langle Z \rangle=2.4$, following Phillips-Thorpe theory\cite{valadez-perez_dynamical_2013}; however, most systems with centro-symmetric interactions, like pure depletion, need to meet the isostaticity criterion from Maxwell counting at $\langle Z \rangle = 6$. In DPD, the addition of complete hydrodynamics allows the network to fully stabilize slightly below this threshold, typically $\langle Z \rangle > 5$; however, in bimodal systems this convention is complicated by the fact that large particles can form significantly higher local $Z$ values when bonding exclusively with small colloid neighbors. Therefore, we do not rely on this definition and instead monitor several additional structural parameters to differentiate between fluids of clusters and percolated gels.

The first of these measures is the size of the largest connected component (LCC) in terms of the total number of particles: $N_{LCC}/N_C$. This value reaches 1 when all particles are fully connected to each other, and is the strictest structural definition of a fully-connected network. Similarly, we track physical percolation of the system by calculating the physical diameter of the LCC as compared to the system box diagonal: $\xi_{LCC}/\sqrt{3L^3}$. When this percolation value is greater than 0.577, the system should span at least 1 axis; at a value greater than 0.816 it should span 2 axes; and, at 1.0 it should span all three. We cross-check this by tracking the physical wrapping of the system across each periodic bound. Monitoring these features over time, we expect that both the LCC and the percolation fraction will plateau around 1.0, when a fully stabilized network has formed.

We also track the structural composition of the other components of the system, those particles that do not belong to the LCC. We calculate the mean size of all other components over time: $\chi_{CC}$. As a system transitions from a fluid to a collection of clusters and finally to a fully percolated gel network, we expect to see the $\chi_{CC}$ parameter rise as separated clusters begin forming, and then peak when these clusters begin merging into a single LCC, falling back toward zero as the system becomes fully connected. Finally, we also evaluate the dynamics of the system using the mean squared displacement of all colloid particles. On a log-log plot, a diffusive system will have a linear slope, and a caged system will plateau as a flat line.

In this framework, a strict gel is therefore defined as having an LCC plateaued at 1.0, a percolation value plateaued at 1.0, a $\chi_{CC}$ parameter that has peaked and decayed to zero, local coordination plateaued at $\langle Z \rangle \geq 6$, and a caged, sub-diffusive MSD. However, there are many real cases where practitioners would consider a gel to have formed without meeting all of these criteria. For our purposes, we define the gel phase to include systems that are clearly forming a percolated space-spanning network, even if it does not meet all of these criteria yet. A system must have a combination of elevated LCC, percolation, or $\langle Z \rangle$ and decaying dynamics or $\chi_{CC}$. The simulation trajectories of all edge cases are also visually inspected to qualitatively confirm the correct phase assignment.

\section*{Results and discussion}

\begin{figure*} [hp!]
    \centering
    \includegraphics[width=1\textwidth]{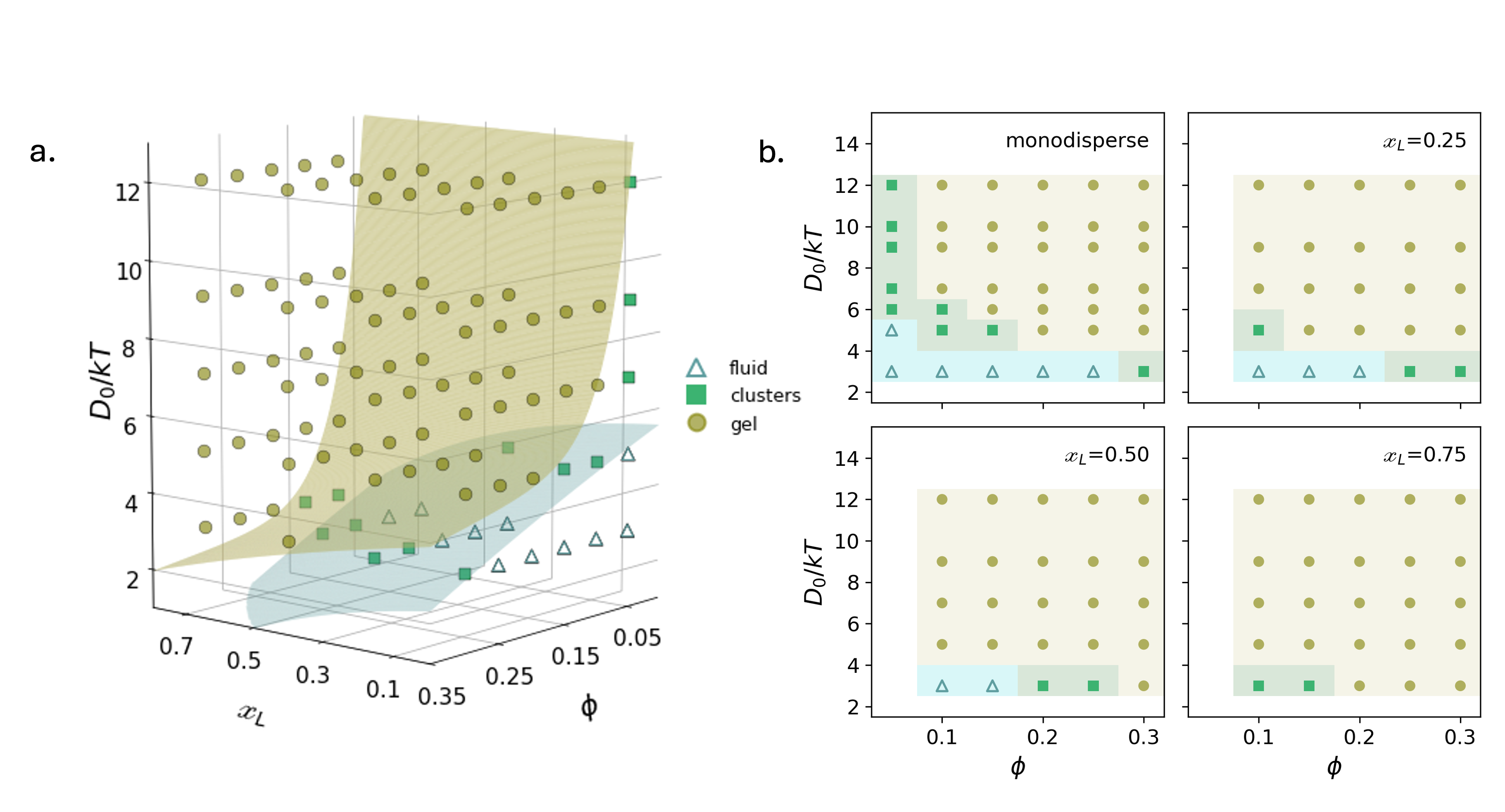} 
    \caption{\textbf{Phase diagram of attractive, bimodal depletion gels.} The total phase space is defined with respect to depletion attraction strength $D_0/kT$, total volume fraction $\phi$, and large particle fraction $x_L$. Markers and color indicate the phase of a given simulation: fluid (blue, open, triangles), clusters (green, filled, squares), or gel (brown, filled, circles). (a) The 3D phase diagram for $D_0/kT=[3,5,7,9,12]$, $\phi=[0.05,0.1,0.15,0.2,0.25,0.3]$ and $x_L=[0.0, 0.25, 0.5, 0.75]$. For bimodal compositions, $\phi=0.05$ is excluded. (b) 2D projections of the 3D diagram for $D_0/kT=[3,5,7,9,12]$ and $\phi=[0.05,0.1,0.15,0.2,0.25,0.3]$ at different $x_L$ values: the monodisperse systems at both $x_L=0$ and $x_L=1.0$ (top left). In this panel only $D_0$ is reported in terms of the local particle size ($D_0^L=a_L\cdot D_0=2D_0$) to make the two monodisperse systems directly comparable to each other. The other panels use a consistent $D_0$ value, as defined in the Methods, and represent data for bimodal compositions at $x_L=0.25$ (top right), $x_L=0.5$ (bottom left), and $x_L=0.75$ (bottom right).}
    \label{fig:3D-phase-diagram}
\end{figure*}

\begin{figure*} [htp!]
    \centering
    \includegraphics[width=1\textwidth]{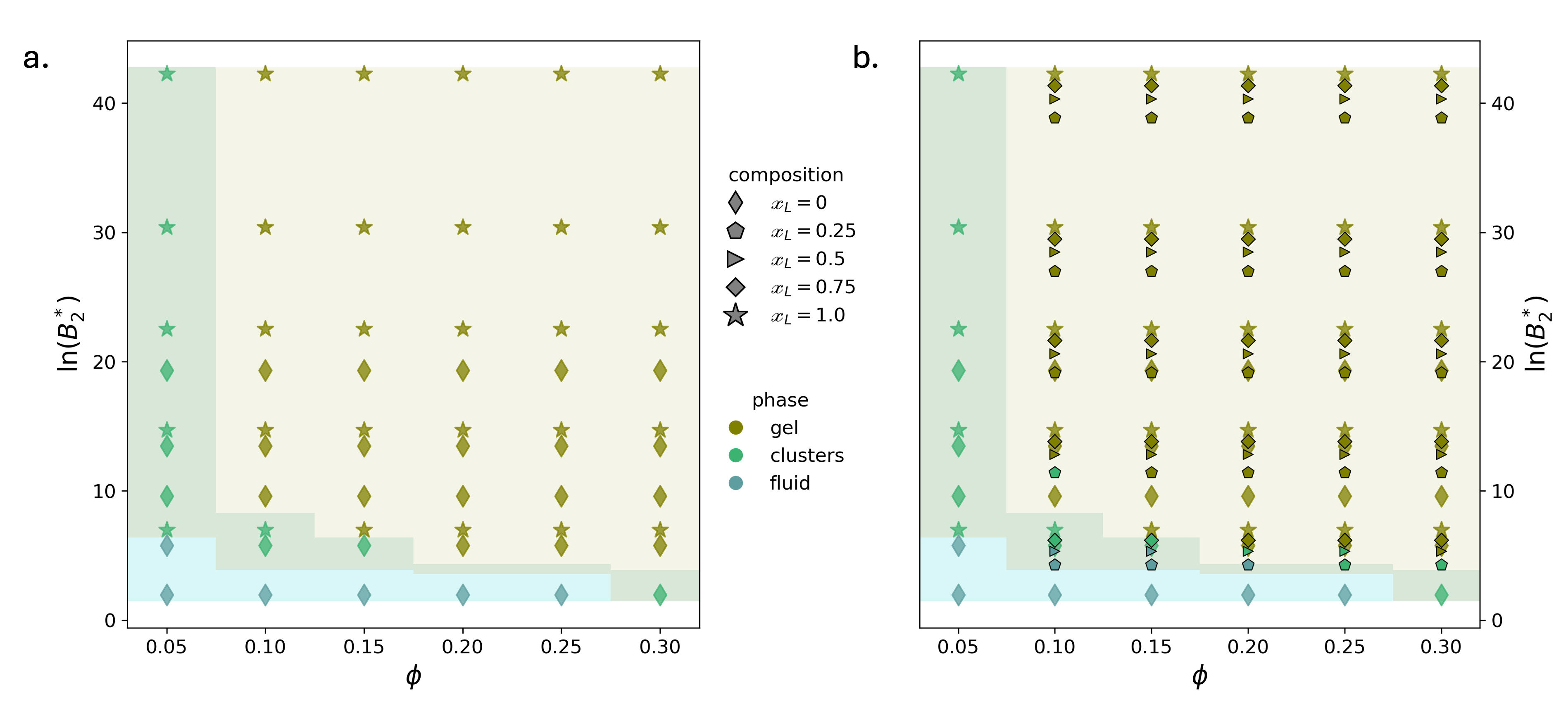} 
    \caption{\textbf{Contact-weighted reduced second virial coefficient collapses bimodal gelation to the monodisperse phase diagram} (a) The monodisperse colloidal depletion gel phase diagram in terms of the log-scaled, contact-weighted reduced second virial coefficient $\ln(B_2^*)$. Markers indicate particle composition: small-colloid gels (narrow diamonds) and large-colloid gels (stars). The fluid phase is labeled in blue, the cluster phase is labeled in green, and the gel phase is labeled in brown. The effective region of each phase is also shaded in the same tone on the diagram (b) The same monodisperse data, overlaid with bimodal systems sorted by their reduced second virial coefficient. The same colors and markers apply. Additional markers indicate the bimodal composition in terms of large particle fraction, $x_L$: 0.25 (pentagons), 0.5 (right-facing triangles), and 0.75 (wide diamonds).}
    \label{fig:2D-phase-diagram}
\end{figure*}

We study the bimodal phase space for colloidal depletion systems by varying three parameters: the total particle volume fraction, $\phi$; the fraction of large particles, $x_L=\phi_L/\phi$, which is defined by the ratio of the volume fraction of large particles to the total volume fraction of the system; and the interparticle attraction, $D_0/kT$, which represents changes to the concentration of depletant. All systems maintain a fixed attraction range consistent with constant depletant size, and are classified as either a monomeric fluid, a fluid of clusters, or a gel, as described in the Methods section. The full phase space, spanning $\phi=0.05-0.3$, $D_0/kT=3-12$, and $x_L=0-1$ is shown in Fig. \ref{fig:3D-phase-diagram}. We find notable variation in system aggregation between the monodisperse and bimodal compositions. Fig. \ref{fig:3D-phase-diagram}a shows the 3D phase space created by varying $x_L=[0.0,0.25,0.5,0.75]$, and Fig. \ref{fig:3D-phase-diagram}b compares the 2D projection of the combined monodisperse case (both $x_L=0.0$ and $x_L=1.0$) with the 2D projections of the three bimodal compositions of $x_L=[0.25,0.5,0.75]$. The monodisperse case comfortably spans all three phases, transitioning from fluid to clusters to gel as either $D_0/kT$ or $\phi$ are increased. At very low volume fraction ($\phi=0.05$) we do not see full percolation of the system, even at the highest attraction, and this volume fraction is excluded from the bimodal case. In the remaining region, we find that colloid aggregation is strongly affected by bimodal composition. Increasing the fraction of large particles in the system shifts the gel line to both lower total volume fraction and weaker relative interactions. Therefore, it appears that the presence of more large particles makes it easier to form a gel. The fact that this enhanced gelation occurs at low volume fraction ($\phi=0.1$) suggests that this effect is not driven by caging, and likely comes from changes in interparticle bond formation. Because large particles have both stronger average attraction and lower average mobility, it is difficult to isolate whether the change in aggregation is a diffusion or reaction limited effect.

We investigate this effect further by calculating the reduced second virial coefficient in terms of the hard-sphere excluded volume, as defined by Noro and Frenkel\cite{NoroFrenkel2000} as $B_{2,ij}^*=B_{2,ij}/B_{2,ij}^{HS}$. These terms are calculated directly from the Morse potential, $U^M_{ij}(r)$ and the hard-sphere diameter $r_0=a_i+a_j$, as:

\begin{equation}
    B_{2,ij} =2\pi \int_{0}^{\inf} \big( 1-e^{-U(r_{ij})/kT}\big) r^2 dr
\label{eq:weighted-second-virial-coeff}
\end{equation}
\begin{equation}
    B_{2,ij}^{HS} = \frac{2\pi}{3} r_{0}^3
\label{eq:second-virial-coeff}
\end{equation}
\hspace{3cm} \\
We emphasize the contact-driven behavior of these systems by calculating a single $B_2^*$ value for each composition using the contact weighted number averages $W_{ii}=\text{X}_i\text{X}_i(2a_{i})^2$ for like pairs and $W_{ij}=2\text{X}_i\text{X}_j(a_{i}+a_{j})^2$ for unlike pairs, where the number fraction of each population is defined as $\text{X}_i=N_{i}/N_{C}$, and the system-specific value becomes:

\begin{equation}
    B_{2}^*=\frac{(B_{2,SS} \cdot W_{SS})+(B_{2,SL} \cdot W_{SL})+(B_{2,LL} \cdot W_{LL})}{(B_{2,SS}^{HS} \cdot W_{SS})+(B_{2,SL}^{HS} \cdot W_{SL})+(B_{2,LL}^{HS} \cdot W_{LL})}
\label{eq:contact-weighted}
\end{equation}
\hspace{3cm} \\
To account for the narrow attraction range used in these simulations, and to more clearly resolve variation within the higher-attraction gel regimes, we compare systems using the natural logarithm of this parameter, $\text{ln}(B_2^*)$. We find that plotting the log of this contact-weighted reduced second virial coefficient versus total volume fraction for all systems combines the effects of composition and attraction into a single axis. Fig. \ref{fig:2D-phase-diagram}a shows the resulting 2D phase diagram for monodisperse data, including both $x_L=0.0$ and $x_L=1.0$. Fig. \ref{fig:2D-phase-diagram}b shows the same data overlaid with the bimodal compositions. As expected the monodisperse phase diagram is equivalent to the $D_0/kT$ vs. $\phi$ plot in Fig. \ref{fig:3D-phase-diagram}b. Additionally, in all but one case the bimodal data now follows the same phase boundaries. There is one outlier at low volume fraction ($\phi=0.1$) and moderate attraction. The monodisperse system at $\text{ln}(B_2^*)=9.60$ (with $x_L=0.0$ and $D_0=7kT$, 8,189 colloids) forms a fully percolated network with $\langle Z \rangle=6.4$, but the bimodal system at slightly higher $\text{ln}(B_2^*)=11.42$ (with $x_L=0.25$ and $D_0=5kT$, 6,397 colloids (6,141 small, 256 large)) achieves less than 10\% connectivity with $\langle Z \rangle=3.3$, placing it in the cluster regime even after $72.4\tau^D_{\langle a_C \rangle}$. This difference in assembly between two systems with equivalent total volume fraction, similar attraction conditions, and similar total number of particles suggests that aggregation of bimodal systems in the weak and intermediate attraction regimes near the gelation boundary may be more affected by the details of particle dynamics.

\begin{figure*} [hp!]
    \centering
    \includegraphics[width=1\textwidth]{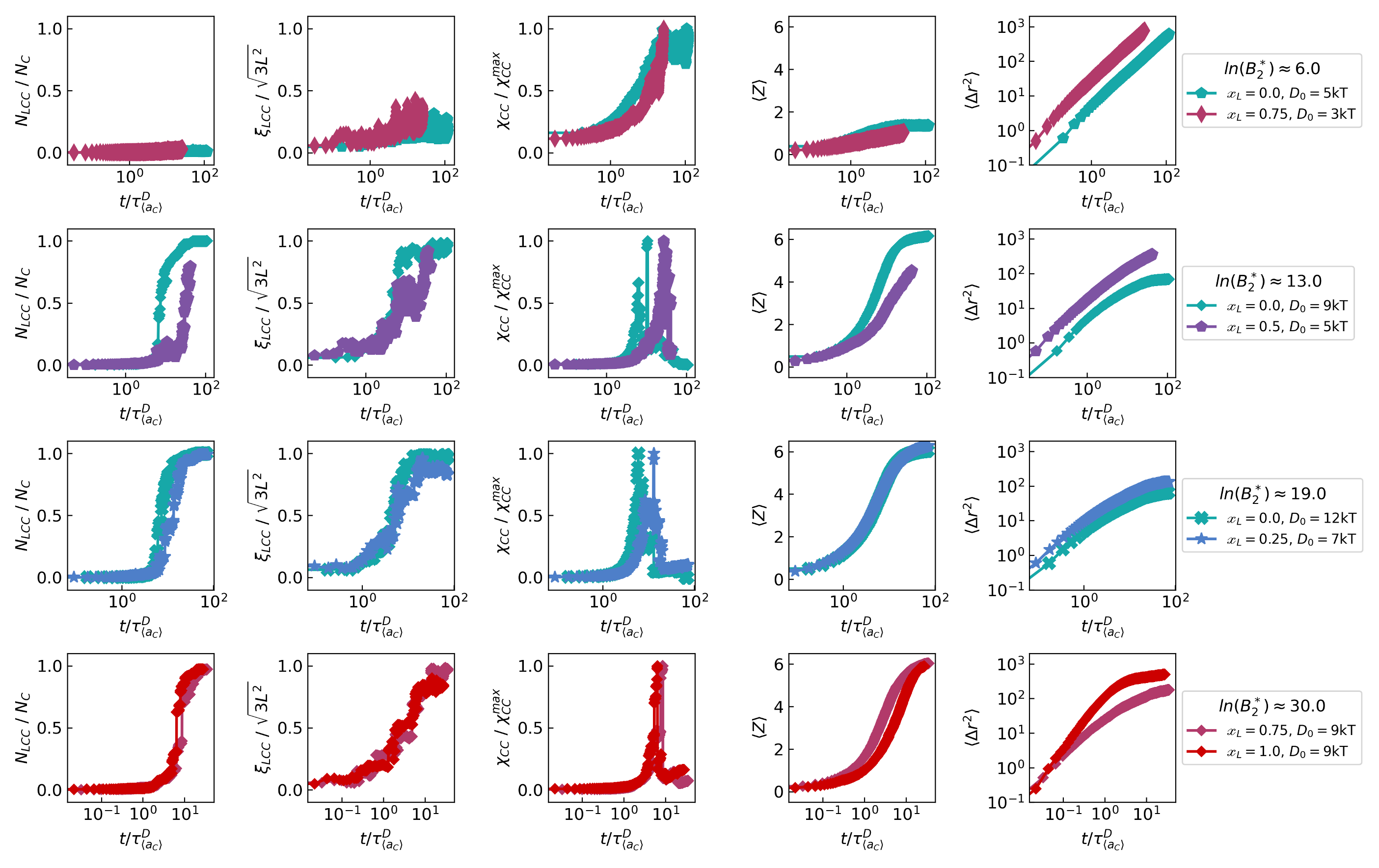} 
    \caption{\textbf{Aggregation dynamics of bimodal depletion gels.} Comparison of aggregation metrics in monodisperse and bimodal colloidal gels at total volume fraction $\phi=0.10$ and matched $ln(B_2^*)$. From left to right, columns represent: (1) the formation of the largest connected component (LCC) as a fraction of total particles $N_{LCC}/N_C$, (2) the percolation of that network, measured as the physical diameter of the LCC divided by the system box diagonal $\xi_{LCC}/\sqrt{3L^3}$, where $L$ is the length of one side of the simulation box, (3) the mean size of all remaining, non-percolated clusters in the system, $\chi_{CC}$, reported as a fraction of the maximum size during the entire simulation $\chi_{CC}^{max}$, (4) the evolution of the average coordination number, $\langle Z \rangle$, and (5) the colloidal mean squared displacement (MSD). From top to bottom, rows represent different values of $\ln{B_2^*} \approx (6.0, 13.0, 19.0, 30.0)$. Each system is plotted in terms of the bare particle diffusion time for the average sized colloid in that system, $\tau_{\langle a_C \rangle}^D$. Colors indicate the fraction of large particles in the system, $x_L$, at values of $0.0$ (teal), $0.25$ (blue), $0.5$ (purple), $0.75$ (dark magenta), and $1.0$ (red). Symbols indicate the depletion attraction strength for each system, $D_0$, at values of $3kT$ (narrow diamond), $5kT$ (pentagon), $7kT$ (star), $9kT$ (wide diamond), and $12kT$ (cross).}
    \label{fig:dynamics_phi10}
\end{figure*}

However, when examining the dynamics of these systems we find that the $\text{ln}(B_2^*)$ mapping also successfully describes their evolution through time in the majority of cases. In Fig. \ref{fig:dynamics_phi10} we compare five dynamical parameters at matched $\text{ln}(B_2^*)$ values and a total volume fraction of $\phi=0.10$. From left to right, the first two columns show the formation of the largest connected component (LCC) and the percolation of that network across the system box diagonal. These measures monitor the percolation transition. Together they describe what fraction of particles are in the main network and if that network physically spans the dimensions of the simulation box. The third column is the mean size of all remaining, non-percolated clusters in the system, $\chi_{CC}$. It is a complementary analysis to percolation that described the aggregation of the rest of the system. During gelation, $\chi_{CC}$ should grow as particles aggregate into clusters before coming to a peak, after which large clusters begin to merge into the LCC and $\chi_{CC}$ decays back to zero. The fourth column plots the evolution of the average coordination number within the entire system, $\langle Z \rangle$. This monitors the internal structure of clusters and the network as they form. A system that percolates at low $\langle Z \rangle\approx 2$ is extremely tenuous, while a more densely packed system with $\langle Z \rangle \approx 6$ meets the isostaticity criterion from Maxwell counting and should produce a bulk elastic modulus. Finally, the last column shows the mean squared displacement (MSD), which monitors whether or not particle motion has become sub-diffusive and the network is stable. 

Across all compositions we find that the majority of these behaviors collapse onto a single curve describing both monodisperse and bimodal aggregation when plotted in terms of $\text{ln}(B_2^*)$ and the bare particle diffusion time of the average-sized particle in the system, $\tau^D_{\langle a_C \rangle}= 6\pi\eta_0 a_C^3/ k{T}$. Like the $\text{ln}(B_2^*)$ phase diagram, this result suggests that aggregation in bimodal systems can be accurately predicted from an equivalent monodisperse representation; however, there are two exceptions. At the intermediate attraction of $\text{ln}(B_2^*)\approx13$, the bimodal system displays delayed aggregation with reduced average coordination number, diffusive dynamics, and delayed percolation of the network. Conversely, at the strong attraction regime of $\text{ln}(B_2^*)=30$ the structural evolution is identical, but the dynamics of the monodisperse system are diffusive for longer, before undergoing a much sharper arrest. Similar deviations are visible at these conditions in all the total volume fractions simulated in this study ($\phi$=[0.10, 0.15, 0.20, 0.25, 0.30]).

\begin{figure*} [hpt!]
    \centering
    \includegraphics[width=1\textwidth]{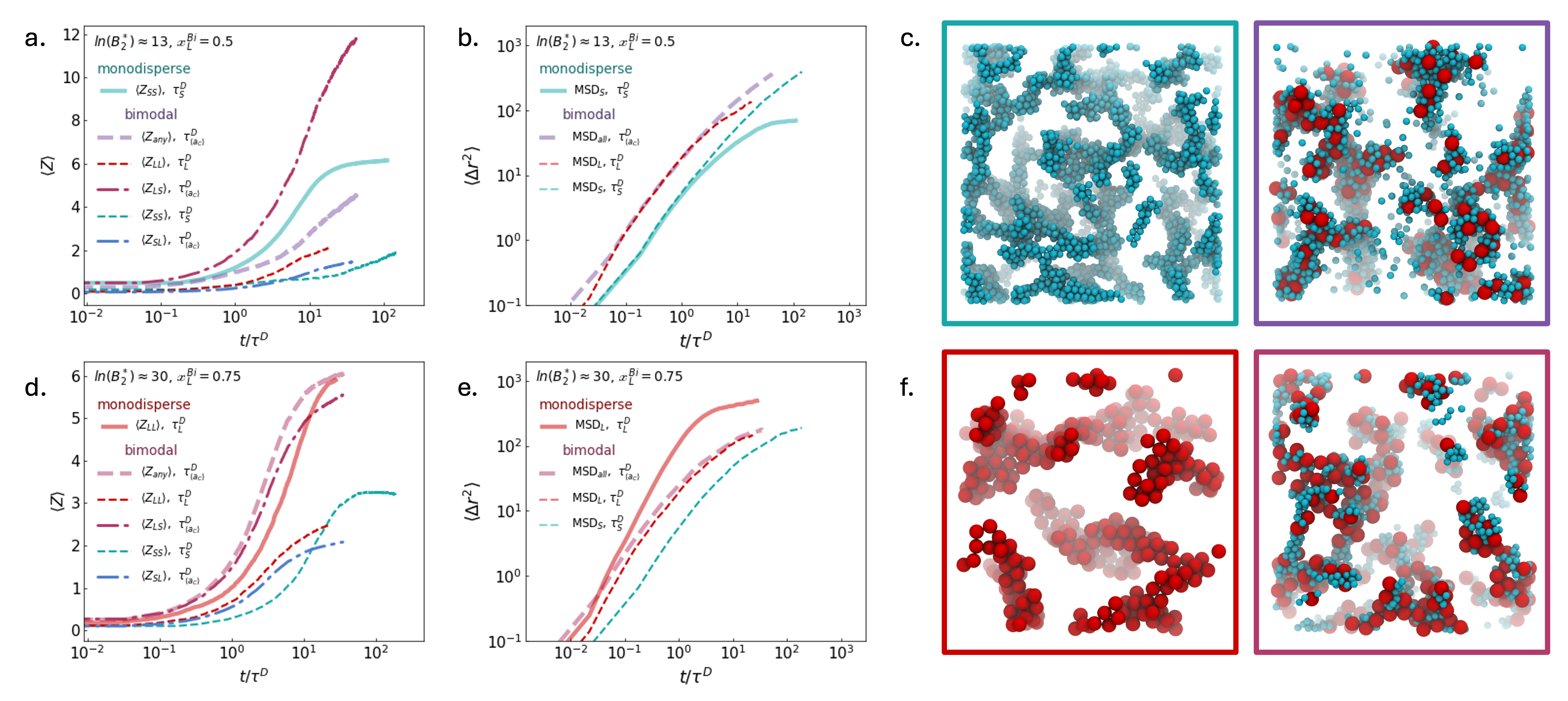} 
    \caption{\textbf{Per-population decomposition of particle dynamics.} All data is for $\phi=0.1$. (a) Decomposition of the average coordination number $\langle Z_{any} \rangle$ (light purple, bold, dashed) of the bimodal system at intermediate attraction of $\ln(B_2^*)=13.0$that was reported in Fig. \ref{fig:dynamics_phi10}. $\langle Z_{any} \rangle$ is decomposed into small-small $\langle Z_{SS} \rangle$ (teal, dashed), small-large $\langle Z_{SL} \rangle$ (blue, dash-dot), large-small $\langle Z_{LS} \rangle$ (maroon, dash-dot), and large-large $\langle Z_{LL} \rangle$ (light red, dashed) interactions. These data are compared to the corresponding monodisperse system's $\langle Z_{SS} \rangle$ (light teal, bold, solid) that was also reported in Fig. \ref{fig:dynamics_phi10}. (b) Decomposition of the mean squared displacement $\text{MSD}_{all}$ (light purple, bold, dashed) for the bimodal system at $\ln(B_2^*)=13.0$ that was also reported in Fig. \ref{fig:dynamics_phi10}. $\text{MSD}_{all}$ is decomposed into the mobility of small particles, $\text{MSD}_{S}$ (teal, dashed) and large particles $\text{MSD}_{L}$ (red, dashed). These data are compared to the corresponding monodisperse system's $\text{MSD}_S$ (light teal, bold, solid) that was also reported in Fig. \ref{fig:dynamics_phi10}. (c) Visualization of a representative slab of the monodisperse and bimodal systems at intermediate attraction of $\ln(B_2^*)=13.0$ (small particles, teal; large particles, red) (d) The same decomposition of average coordination number for very-strong attraction of $\ln(B_2^*)=30.0$. $\langle Z_{any} \rangle$ is recolored (light pink, bold, dashed) and the monodisperse system is labeled $\langle Z_{LL} \rangle$ (light red, bold, solid) to match Fig. \ref{fig:dynamics_phi10}. (e) The same decomposition of MSD for $\ln(B_2^*)=30.0$. $\text{MSD}_{all}$ is recolored to match $\langle Z_{any} \rangle$ in d, and the monodisperse data is relabeled $\text{MSD}_{LL}$ and recolored to match $\langle Z_{LL} \rangle$ in d. (f) Visualization of a representative slab of the monodisperse and bimodal data at very-strong attraction of $\ln(B_2^*)=30.0$.}
    \label{fig:per-pop-dynamics}
\end{figure*}

These deviations in dynamics appear to be driven by the preferential formation of small-large particle bonds. Fig. \ref{fig:per-pop-dynamics} shows the per-population decomposition of average coordination number and mean square displacement for pairs in the $\phi=0.1$ systems at matched $\text{ln}(B_2^*)$ conditions. In these plots the bold, transparent, solid line represents the monodisperse dataset and the bold, transparent, dashed line represents the average behavior of the bimodal dataset. These are the same data that were plotted in Fig. \ref{fig:dynamics_phi10}. We decompose the bimodal system by population and consider small-small, small-large, large-small, and large-large contacts separately. For the directional coordination numbers, $\langle Z_{SL} \rangle$ is defined as the number of large particles connected to the same central small particle, and $\langle Z_{LS} \rangle$ is defined as the opposite relationship, the number of small particles connected to the same central large particle. Comparing these values to the monodisperse case, we see that the bimodal system is divided into two structural regimes. Fig. \ref{fig:per-pop-dynamics}a shows that, at an intermediate attraction of $\text{ln}(B_2^*)=13$, the bimodal system achieves a lower total $\langle Z \rangle \approx 4$ compared to the monodisperse system, which reaches $\langle Z \rangle\approx 6$. In the bimodal system, the $\langle Z \rangle$ also rises at a slower rate. This reduced aggregation is explained by the small-small, large-large, and small-large bonds, which are all significantly suppressed below $\langle Z \rangle=2$. In sharp contrast, the large-small contacts are significantly enhanced, reaching $\langle Z \rangle \approx 12$. This is because large particles have stronger attraction and around 4$\times$ the surface area of a small particle, allowing them to accumulate many more neighbors. At bimodal composition $x_L=0.5$ there are also a considerable number of large particles in this system, suggesting that the total percolation of this network depends heavily on these large-small coordination hubs. Although the total average coordination number seems to accurately describe the phase behavior of the system, it fails to account for this structural heterogeneity. Looking at the same decomposition for the very-strong attraction condition of $\text{ln}(B_2^*)=30$ in Fig. \ref{fig:dynamics_phi10}d we find the same type of structural heterogeneity. Here the monodisperse system and the bimodal system aligned in Fig. \ref{fig:dynamics_phi10}, but the decomposition makes it clear that this is because, at bimodal composition $x_L=0.75$, the large-small contacts dominate the structure. Their average coordination number is not nearly as enhanced as in the intermediate attraction case, but again it is significantly higher than the small-small, small-large, and large-large contacts. This type of structural heterogeneity does not exist in a monodisperse system and therefore cannot be predicted from an equivalent monodisperse condition.

Fig. \ref{fig:dynamics_phi10}b shows the decomposed MSD for the intermediate attraction case. Again the bold, transparent solid line is the monodisperse dataset and the bold, transparent, dashed line is the average behavior of the bimodal dataset, the same data that was plotted in Fig. \ref{fig:dynamics_phi10}. We decompose the MSD by tracking per-particle motion and recalculating population specific MSD values. Here the value of $\tau_D$ changes with the population and always reflects the diffusion time of the average sized particle under consideration (i.e. $\tau^D_S$ for small particles, $\tau^D_L$ for large particles, and $\tau^D_{\langle a_C \rangle}$ for the full bimodal population). Comparing the bimodal $\text{MSD}_S$ with the small-particle monodisperse data we see that small particles originally follow the same behavior in both bimodal and monodisperse systems. However, at later times when the monodisperse system is arresting, the small particles in the bimodal system remain diffusive. Instead the total bimodal behavior is primarily described by $\text{MSD}_L$, the dynamics of the large particles. This reinforces the centrality of large particles to the network structure. The fact that small particles show long-time diffusive behavior in this system despite the formation of dense large-particle based coordination hubs (reaching $\langle Z_{LS} \rangle \approx 12$) also suggests that the network is less spatially homogeneous than the monodisperse condition, and exhibits a larger void size. This would make it harder for disconnected small particles to find contacts at later times, prolonging their diffusivity and delaying the total percolation of the network. The very-strong attraction case in Fig. \ref{fig:dynamics_phi10}e shows the opposite trend. Here the bimodal $\text{MSD}_L$ quickly diverges from the large-particle monodisperse case and more closely follows the bimodal $\text{MSD}_S$. This behavior contributes to the earlier and more gradual arrest of the bulk. At $x_L=0.75$, this suggests that the higher attraction of large particles allows them to accumulate small neighbors more quickly, caging their dynamics and facilitating percolation of the network through the use of small contacts. These results show that even when the phase behavior of a bimodal system can be described through mean-field information, the underlying network structure may still depend on composition and population specific effects related to both attraction and particle mobility.

\begin{figure*} [ht]
    \centering
    \includegraphics[width=0.9\textwidth]{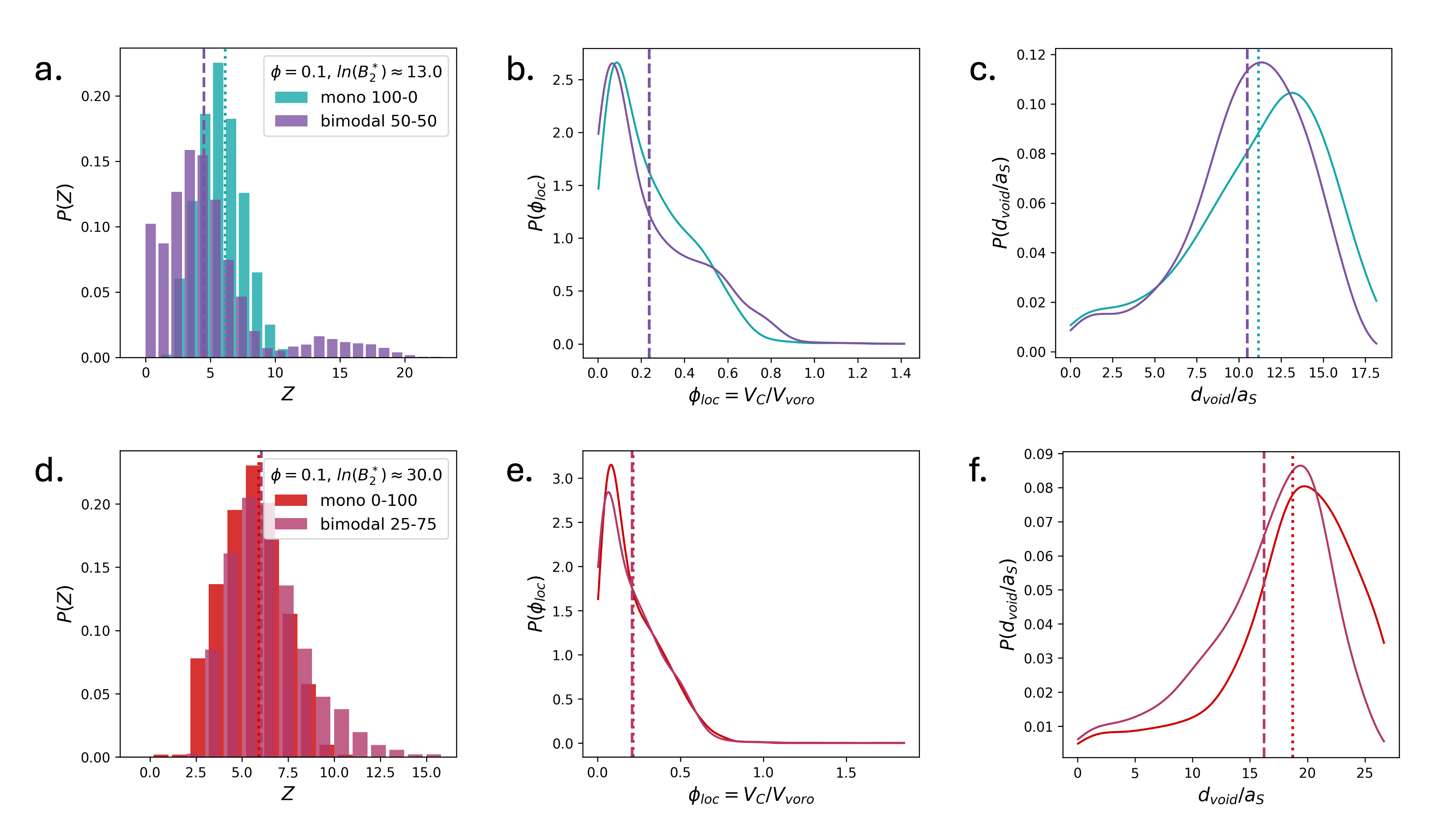} 
    \caption{\textbf{Variation in structure distributions at $\phi=0.1$}. Vertical lines mark the distribution mean for monodisperse (dotted) and bimodal (dashed) systems. (a) The distribution of local coordination number $Z$ for matched monodisperse (teal) and bimodal (purple) systems at $\ln(B_2^*)=13.0$. (b) The same comparison for the local packing fraction $\phi_{loc}=V_C/V_{voro}$. (c) The same comparison for void diameter $d_{void}/a_S$. (d) The distribution of local coordination number $Z$ for monodisperse (red) and bimodal (dark magenta) systems matched at $\ln(B_2^*)=30.0$. (e) The same comparison for local packing fraction. (f) The same comparison for void diameter.}
    \label{fig:structure-distributions-phi10}
\end{figure*}

\begin{figure*} [htp!]
    \centering
    \includegraphics[width=0.9\textwidth]{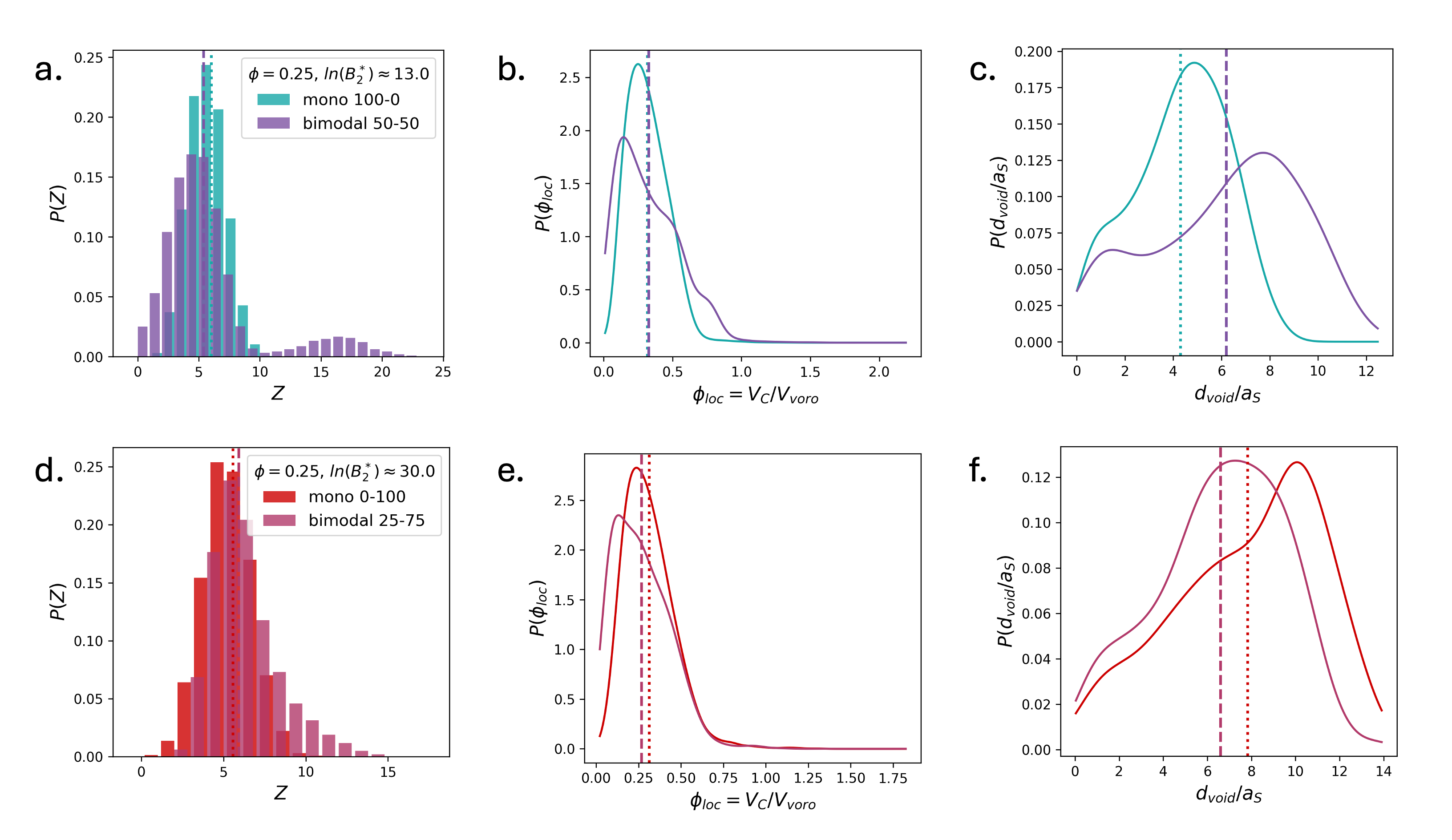} 
    \caption{\textbf{Variation in structure distributions at $\phi=0.25$}. Vertical lines mark the distribution mean for monodisperse (dotted) and bimodal (dashed) systems. (a) The distribution of local coordination number $Z$ for matched monodisperse (teal) and bimodal (purple) systems at $\ln(B_2^*)=13.0$. (b) The same comparison for the local packing fraction $\phi_{loc}=V_C/V_{voro}$. (c) The same comparison for void diameter $d_{void}/a_S$. (d) The distribution of local coordination number $Z$ for monodisperse (red) and bimodal (dark magenta) systems matched at $\ln(B_2^*)=30.0$. (e) The same comparison for local packing fraction. (f) The same comparison for void diameter.}
    \label{fig:structure-distributions-phi25}
\end{figure*}

This is further demonstrated by visual inspection of the systems and a comparison of their underlying depletion conditions. Fig. \ref{fig:per-pop-dynamics}c visualizes a representative slab of the final snapshot of the two intermediate attraction cases. It is clear that the bimodal system has not fully gelled but does appear to be forming a space spanning network. As evidenced by the $\langle Z \rangle$ and MSD data, it contains dense hubs of large particles decorated by many small particles that link together into the backbone of this network. This gives the bimodal gel a qualitatively distinct structure from the monodisperse gel. The bimodal network is less homogeneously distributed through the simulation box and has larger voids. Fig. \ref{fig:per-pop-dynamics}f visualizes a representative slab of the final snapshot of the two very-strong attraction cases. Here both the monodisperse and bimodal structures are fully arrested and appear more structurally similar. As evidenced by the $\langle Z \rangle$ data, this bimodal structure is more segregated than at the intermediate condition with small particles forming extended small-only regions that bridge between regions of large particles. This type of structure can be partially explained by the details of the depletion conditions. In the very-strong regime, the monodisperse and bimodal systems share the same $\text{ln}(B_2^*)$ and the same depletion attraction of $D_0=9kT$. In the large monodisperse system this produces $U_0^{LL}=18kT$, and in the bimodal system this produces an attraction hierarchy of $U_0^{SS}=9kT$, $U_0^{SL}=13.5kT$, and $U_0^{LL}=18kT$. All of these conditions fall well within the gelation regime shown in Fig. \ref{fig:3D-phase-diagram}b. This suggests that mixing them together facilitates clustering and caging of all particle types, producing the earlier and more gradual arrest shown in Fig. \ref{fig:dynamics_phi10}e. In contrast, in the intermediate attraction case the small-particle monodisperse system is at $D_0=9kT$ but the bimodal system is only at $D_0=5kT$, which is close to the gelation boundary. This produces a bimodal attraction hierarchy that extends into the gelation regime: $U_0^{SS}=5kT$, $U_0^{SL}=7.5kT$, and $U_0^{LL}=10kT$. Nevertheless, at $x_L=0.75$ a quarter of the particle volume is still made of small particles that experience weaker attraction. As a result, even though the bimodal system averages $\ln(B_2^*)=13$ it behaves more like a mixture of two separate phases (a weakly attractive gel and a strongly attractive gel), resulting in a distinct assembly pathway and structure. Both the intermediate and very-strong attraction conditions highlight how the use of mean-field approximations like the contact-weighted reduced second virial coefficient can obscure details of bimodal structure and dynamics.

This limitation is important because the structural details of a bimodal system likely influence its bulk mechanics. Previous studies have shown that mechanical properties can vary widely at the same $\langle Z \rangle$ \cite{Rocklin2021}, and that bulk mechanics are often more reliably predicted from mesoscale structure \cite{Whitaker2019,Nabizadeh2024}. To quantitatively evaluate how much the bimodal structure diverges from the matched monodisperse condition we examine the distribution of coordination number, local packing fraction, and void size in both the intermediate and very-strong attraction regimes at $\phi=0.1$. These distributions are shown in Fig. \ref{fig:structure-distributions-phi10}. The local packing fraction is defined per-particle as the particle volume divided by its Voronoi volume, $\phi_{loc}=(4/3)\pi a_C^3/V_{voro}$. The mesoscale void size is calculated from a random sample of 10,000 points using the Gubbins Pore Size Distribution, where each point is defined as the center of the largest spherical void that fits inside the gel at that location \cite{Sorichetti2020,GelbGubbins1999}. While the mean values for all three parameters are fairly consistent, the distributions show more variation. At the intermediate condition, Fig. \ref{fig:structure-distributions-phi10}a,b shows that the coordination number and the local packing fraction in the bimodal system (purple) both show evidence of a bimodal distribution. This is consistent with the existence of two populations of particles with distinct microscale interactions. In contrast, at the very-strong condition Fig. \ref{fig:structure-distributions-phi10}d,e shows that the bimodal system remains more monomodal (dark magenta) and has a similar structure to the equivalent monodisperse system. At both attraction conditions there is less variation in the void size between monodisperse and bimodal gels (Fig. \ref{fig:structure-distributions-phi10}c,f); however, this changes if we increase the total volume fraction to $\phi=0.25$. Here, the bimodal intermediate condition fully gels, and Fig. \ref{fig:structure-distributions-phi25}a shows that this produced a more cleanly bimodal coordination number distribution (purple). In Fig. \ref{fig:structure-distributions-phi25}b we show that the local packing fraction for the bimodal system has also become more heterogeneous (purple), and Fig. \ref{fig:structure-distributions-phi25}c now shows a wide, bimodal distribution of void sizes (purple). These results reflect a significantly more heterogeneous gel network overall, suggesting that bimodal effects can propagate across structural length scales and tend to become more significant at higher volume fraction. This conclusion is supported by the data for the very-strong attraction condition. Fig. \ref{fig:structure-distributions-phi25}d shows that the coordination number distribution remains similar in both the bimodal and monodisperse gels; however, the distribution of local packing fraction and void size (Fig. \ref{fig:structure-distributions-phi25}e,f, dark magenta) both clearly diverge from the monodisperse case.

\section*{Conclusions}

In summary, we find that the stoichiometric composition of attractive, bimodal colloidal depletion gels alters their assembly and gelation. At fixed total volume fraction, increasing the fraction of large particles shifts the gel boundary to lower volume fractions and weaker attraction strengths. This change in the gel boundary is qualitatively similar to the effect of adding non-interactive Brownian inclusions, where a decrease in available free volume increases aggregation; however, in a fully attractive system the change in phase behavior appears to be driven by the existence of heterotypic attractive interactions. We find that the bulk phase behavior of the bimodal systems can be well described by combining interaction strength and bimodal composition into one contact-weighted reduced second virial coefficient. This attractive parameter and the total volume fraction appear to describe the majority of the gelation phase diagram. Additionally, this approach is successful at describing a large portion of bulk dynamics and structural assembly. Across a wide range of values $\ln(B_2^*)=[6,13,19,30]$, the majority of particle aggregation and mobility during the gelation process can be described in terms of the contact-weighted second virial coefficient and the bare particle diffusion time for the average sized particle. However, inconsistencies in the phase behavior between bimodal and monodisperse systems near the gel line at low total volume fraction ($\phi=0.10$) point to broader bimodal effects on colloidal gelation. 

Across all volume fractions, the dynamics and assembly of bimodal systems at both intermediate attraction and very-strong attraction diverge from their matched monodisperse cases. By decomposing the mobility and the average coordination number by particle population and contact type, we show that the bimodal systems are defined by heterotypic small-large interactions and inter-species variation in attraction strength. Adding large particles appears to increase the bonding capacity of the system by locally altering mobility, the attraction environment, and in some cases phase behavior. Depletion systems exhibit an implicit attraction hierarchy that changes the interaction strength between different bimodal pairs in the same system. Within the same hierarchy range, ex: size ratio 1:2, such that $U_0^{SL}=1.5U_0^{SS}$ and $U_0^{LL}=2U_0^{SS}$, the effect of bimodality varies depending on which portion of the phase diagram this hierarchy spans. In the very-strong case, when the hierarchy is firmly in the gelation region, bimodality smooths and accelerates arrest. This produces a network with a similar distribution of local contacts to the monodisperse case. However, in the intermediate case, when the hierarchy includes the region near the gel boundary and the fully gelled regime, the local particle interactions are highly heterogeneous. Large particles form densely connected coordination hubs that are connected by more weakly attractive strands of small particles. This system displays delayed and far more tenuous aggregation, as compared to a monodisperse system in the same attraction regime.

We also show that these effects can propagate across length scales in the network, altering mesoscale structure. At low total volume fraction bimodal effects are largely restricted to the microscale, where coordination number distribution and local packing fraction take on bimodal distributions in the intermediate attraction regime. However, as the volume fraction increases, these bimodal distributions propagate to the mesoscale, altering the distribution of voids in the network. More surprisingly, distortions to local packing fraction and mesoscale structure also emerge in the very-strong case, where local packing is nearly identical to the monodisperse gel. This suggests that bimodality can alter larger length scale structures, even when the effects on microscale packing are minimal. This has considerable implications for the mechanical properties of these systems, which typically emerge through hierarchical interactions across multiscale network structures. Therefore, although the bulk behavior of an attractive, bimodal depletion gel can be largely described by the monodisperse model using mean-field approximations, these mean-field measurements obscure many effects of bimodality on mechanically relevant multiscale structure. This suggests that bimodal composition may be a useful parameter for modifying network features within a consistent aggregation and percolation regime.

\section*{Author contributions}
S.J. and A.M. supervised the research. R.A.C. performed all numerical simulations and data analysis and drafted the manuscript. All authors discussed the results and contributed to the final conclusions and participated in review and editing of the final manuscript.

\section*{Conflicts of interest}
There are no conflicts to declare


\section*{Acknowledgements}
Financial support for this study was provided to S.J. and A.M. by the National Science Foundation (PMP-2025613), and to R.A.C. and S.J. by NASA ROSE FINESST (80NSSC23K0015). Computational resources were provided by the Massachusetts Green High-Performance Computing Center in Holyoke, MA.



\balance

\renewcommand\refname{References}

\bibliography{references} 
\bibliographystyle{rsc} 

\end{document}